# DISCUSSION OF: STATISTICAL ANALYSIS OF AN ARCHEOLOGICAL FIND

By Camil Fuchs

*Tel Aviv University*

**1. Introduction.** The starting points of Feuerverger's paper are both exciting and promising: A scientific puzzle of major importance is settled by a novel statistical approach. The puzzle is related to the re-analyzed inscriptions on the ossuaries from an ancient tomb from Jerusalem unearthed in 1980. The new analysis, also documented in a book [Jacobovici and Pellegrino (2007)] and a documentary movie [Cameron (2007)], claims that the inscriptions indicate that this may be the burial site of the New Testament (NT) family. Undoubtedly, if validated, a discovery with potential to stir major interest both in academic as well as in religious circles. At this point, the statistical methodology is called to settle the controversy and a new statistical approach is developed to handle the intricacies of the complex problem.

The results presented in the paper seem to justify the prior excitement. In terms of the new approach, the defined level of "surprisingness" for the cluster of names in the tomb is found to be very high, that is, under the specified provisos, there is a very low probability that a random sample of such ossuaries contains a cluster of names which is more surprising than the cluster found. Furthermore, when the probabilities related to the level of surprisingness are translated into the classical terms of conditional odds ratios, the odds that the Talpiot tomb is that of the NT family are also found to be very high.

It seems like the statistical methodology succeeded in settling the controversy, and the verdict is in favor of the tomb being the NT family tomb. In the process, a new approach was developed to settle cases in which judgment has to be rendered on whether or not a multiple characteristics event is or is not a result of random draws.

On a personal note, I confess that I would have been very pleased to be able to conclude my discussion with two positive statements: (a) that I found the results convincing and we can second Prof. Feuerverger's claim that the tomb is most likely that of the NT family, and (b) that the new approach is preferable to the existing methods in deciding whether the tested object is the special one.

Unfortunately, to anticipate the findings detailed below, despite the initial excitement and the personal preferences, I find myself in disagreement with the results and the conclusions. As for the new approach, it may evolve and prove beneficial, although not necessarily preferable to existing methods. I believe its properties have yet to be investigated.

**2. The statistical analysis.** Let us first briefly review the relevant statistical features in Feuerverger's approach and their application to the particular data set. The justifications of the above-mentioned contentions are presented in this context.

The analogue of a null hypothesis $H_0$ is defined to be the assertion that the observed configuration of names (on the ossuaries in the tombsite) arose by purely random draws from the onomasticon. The alternative $H_1$ is presented as "an opposite of $H_0$ relevant to the "NT hypothesis" that the tombsite is that of the NT family." An intermediate formulation (with weaker $H_1$) is also presented, with $H_0$ being the assertion that all possible tombs comparable to that of Talpiot arouse under random assignment of names and $H_1$ is the event that among the such possible tombs, one *unspecified* tomb is that of the NT family. With respect to the intermediate $H_1$ and for various prior-like probabilities, Feuerverger assesses from the $H_0$-tail area the odds ratios of the event that the Talpiot tombsite is that of the NT family.

The data from the Talpiot tomb includes six inscribed ossuaries with the following inscriptions:
#1:$M\alpha\rho\iota\alpha\mu\eta\nu o\upsilon\ [\eta]\ M\alpha\rho\alpha$, #2: יהודה בר ישוע, #3: מתיה, #4: ישוע בר יוסף, #5: יוסה, #6: מריה
transliterated as:
#1: Mariamene [$\eta$] Mara, #2: Yehuda son of Yeshua, #3: Matya, #4: Yeshua son of Yoseph, #5: Yoseh, #6: Marya.

At least some of the names are reminiscent of the names related to the NT family. As a first step in determining how significant or (in terms of the proposed approach) how "surprising" is this find, one has to assess how common were those names in the vicinity of Jerusalem in the late Second Temple period. Table 1 presents the frequencies and the relative frequencies of the generic names out of the total compiled male and female nonfictitious names from ossuary and non-ossuary sources [Ilan (2002)]. Furthermore, the table also presents the frequencies and relative frequencies of the relevant renditions of Mary/Mariam and Yoseph from ossuary sources.

Under the proposed approach, the data analysis conditions on both the number of inscribed ossuaries and their gender distribution, as well as on the generational sequence in two of the four male ossuaries. However, the basic analysis deals only with the inscriptions from five ossuaries, with the Yehuda son of Yeshua ossuary being discarded.

TABLE 1
*Frequencies of the named inscribed in the Talpiot ossuary*

| | All sources | | | Ossuary sources | |
|---|---|---|---|---|---|
| Generic name | Frequency | Relative frequency | Renditions | Frequency | Relative frequency |
| | | | **Female** | | |
| Mary/Mariam | 74 | 0.233 | Mariamene | 1 | 0.023 |
| | | | Marya | 13 | 0.295 |
| Females—Total all sources | 317 | | Total ossuaries—Females named Mariam | 44 | |
| | | | **Male** | | |
| Yehuda | 171 | 0.068 | | | |
| Yeshua | 101 | 0.040 | | | |
| Matya/Mattityahu | 62 | 0.026 | | | |
| Yoseph | 221 | 0.088 | Yoseh | 7 | 0.152 |
| Males—Total all sources | 2509 | | Total ossuaries—Males named Joseph | 46 | |

Now, the new approach defines "an a priori defined" measure of "surprisingness" related to the $H_0$–$H_1$ continuum. The "surprisingness" value of a particular configuration increases as the configuration is in some respect closer to $H_1$. The reciprocal form of the "surprisingness" value is defined as "relevance and rareness" (RR value). "Relevance" refers to membership in an a priori list of candidates for inclusion in an NT tombsite, and "rareness" is defined relative to an a priori list of nested possible name renditions for each such candidate. The initial relevant lists are supposed to include names which are reasonable to assume that they have potential to be found in a NT family tomb, based on a set of a-priori formulated hypotheses. The relevant lists have to reflect those hypotheses. In addition, the relevant lists are also allowed to include unrelated names, defined as "Other," as possibly belonging to persons about whom there are no records. The population and the sample are stratified, and separate a priori lists of tomb candidate name renditions are compiled by gender.

In the analysis of the Talpiot data the following assumingly a priori lists of tomb candidate name renditions for men and women are presented:
Men: Yoseph, Yeshua, Yoseh, James and "Other"
Women: Mary Magdalene (denoted MM or Mariamene), Marya, Mariam, Salome and "Other"

Thus the Matya from ossuary #3 is considered as "Other" (one of those possibly belonging to persons about whom there are no records), and Mariamene [$\eta$] Mara is added to the women's list as being "the most specific appellation to Mary Magdalene from among those known." As can be seen from Table 1, this is the only such exact rendition of Mariam among the recorded names.

The RR value of a datum or of a subset of data is defined as the adjusted relative frequency of occurrence of the components under independent random sampling from the onomasticon. The RR for a generic name is its relative frequency, while the RR value for a particular rendition of a generic name is computed as a product of the name's overall relative frequency and the relative frequency from ossuaries sources of the particular rendition within the generic name. For some particular configurations, quite complex (and relatively reasonable) definitional adjustments imposed by $H_1$ are used in the computation of the RR values. In particular, a prized bonus is applied when Yoseph is the father and Yeshua is the son with the RR-value being divided by 1.2.

Under the suggested approach, the names defined as "Other" receive an RR value of 1, and thus have no effect on the product which yields the RR value for the entire cluster. As expected, and as illustrated below, a sample's RR value is critically affected by the two major features of the approach: the definition of the a priori list and the value given to names defined as "Other."

Table 2 presents the RR values for the cluster of names found in the Talpiot tombsite. We can see that Matya is assigned an RR of 1, while the ossuary #2 is discarded (with its two names, Yehuda and Yeshua, but the name Yeshua does appear in the table from ossuary #4).

The product of the individual RR-values yields $1.74 \times 10^{-8}$. Following the division by the prized bonus factor of 1.2, the RR-value for the cluster is $1.45 \times 10^{-8}$. Clusters with a similar configuration (i.e,. two female and three male ossuaries, where one male ossuary has two men in father–son generational alignment) and with a lower RR value are considered to be more "surprising" than the studied tombsite. Out of the $n_1$ and $n_2$ male and female persons in the population, the total possible number of such samples is $n_1^4 \cdot n_2^2$ and the total number of valid samples (which pass pre-specified "reality" requirements) is $\beta n_1^4 \cdot n_2^2$ with $\beta < 1$. In this case, $n_1 = 2509$, $n_2 = 317$ and Feuerverger found that $\beta = 0.906$, yielding $\beta n_1^4 \cdot n_2^2 = 1.981 \cdot 10^{12}$. Among them a proportion of $5.89 \times 10^{-7}$, or about $1/1{,}821{,}000$ have an RR value lower than $1.45 \times 10^{-8}$. The size of the estimated population who could have been interred in ossuaries includes about 4,400 males and 2,200 females. Dividing those values into the studied configuration of 4 male and 2 female inscriptions we obtain an estimate of 1,100 potential "trials" with which the Talpiot tombsite has to be compared.

The $p$-value for testing the alternative that among the comparable possible tombs one *unspecified* tomb is that of the NT family is assessed by the probability that at least one among the 1,100 would have an $H_0$-tail area less or equal to $5.89 \times 10^{-7}$. This probability is bounded above by $1/1{,}655$. For the Bayes-type computation of the posterior probability that this is indeed the NT family tombsite, Feuerverger defines by $\theta$ the (prior) probability that an NT family tomb would consist of a cluster of with an RR value as surprising as that at Talpiot. For $\theta = 1, 0.5$ and $0.1$, the posterior probabilities are 0.9994, 0.9988 and 0.9940, respectively.

In a nutshell, the exposition above reviews the basics of the new proposed approach as applied to the specific data set.

**3. The a priori hypotheses.** As emphasized, the foundation for the analyses is a set of "hypotheses, assumptions and conditions upon which the computations are carried out, collectively named as the provisos." The paper presents detailed explanations and justifications for the features in the provisos and performs sensitivity analysis under some variations of the provisos. We concentrate here on the basic provisos and the resulting conclusions. Undoubtedly, the posterior probabilities are impressive and seem to suggest that this is indeed the tombsite of the NT family.

However, are the provisos reasonable? And more importantly, were the provisos specified and were the analyses carried out according to the stated premises of the new approach? And if not, what is the likely effect of the deviations from those premises?

The a priori nature of the provisos is among the most important premises of the new approach. In this context, let us revisit first the issue of the female names contained in the presumably a priori list of candidates.

3.1. *The female names in the a priori list of candidates.* The list of potential candidates includes the names Mariam and Salome "commonly believed to be" Jesus' sisters, Marya (Jesus' mother), and Mary Magdalene. The addition of Mary Magdalene is explained by the fact that Mary Magdalene was "present at the burial ritual." The contention that Mary Magdalene's ossuary is presumed to be that inscribed as Mariamene [η] Mara is justified by stating that Mariamne is "the most specific appellation to Mary Magdalene from among those known." But it is difficult to avoid the feeling that in a truly a priori compiled list, the probability of adding persons whose relation was only that they were "present at the burial ritual" and had no familial relationship, were likely to be quite low. (The issue of possible familial relationship is discussed, but the addition of the name is not based on it.) Moreover, the addition of the particular rendition of the name to the list gives a clear impression that after observing the data, the list was biased in favor of $H_1$.

TABLE 2
*RR-values for the cluster of names in the Talpiot tombsite*

| Ossuary | Name on ossuary | Generic name | All sources Relative frequency | Renditions | Ossuary sources Relative frequency | RR |
|---|---|---|---|---|---|---|
| | | | **Female** | | | |
| #1 | Mariamene | Mary/Mariam | 0.233 | Mariamene | 0.023 | $0.0053 = 0.23 \cdot 0.02$ |
| #6 | Marya | | | Marya | 0.295 | $0.0690 = 0.23 \cdot 0.30$ |
| | | | **Male** | | | |
| #2 | Yehuda | Yehuda | 0.068 | | | Discarded |
| #4 | Yeshua | Yeshua | 0.040 | | | 0.0403 |
| #3 | Matya | Matya/Mattityahu | 0.026 | | | 1 |
| #4 | Yoseph | Yoseph | 0.088 | | | 0.0881 |
| #5 | Yoseh | Yoseph | 0.088 | Yoseh | 0.152 | $0.0134 = 0.09 \cdot 0.15$ |

Furthermore, since the particular rendition is in the relevant list, the inscription Mariamenou [$\eta$] Mara is now presented as being a unique rendition of Mariam both from ossuary as well as from nonossuary sources. The assigned RR value to that name is 1.68/317, with the *largest* effect on the overall RR value. Clearly, if there is evidence that the elegantly rendered ossuary inscribed Mariamenou [$\eta$] Mara is indeed the ossuary of the Mary Magdalene, the finding is sensational by itself. But if we only use the statistical evidence, the fact that the effect on the overall result of the inscription Mariamenou [$\eta$] Mara (whose presence on the list is at least more ambiguous than the other names) is problematic, to say the least. Were Mariamenou [$\eta$] Mara treated as "Other," the overall RR value would have been 188 times higher, with the corresponding effect on the computed *p*-value.

The effect of the inscription Mariamene [$\eta$] Mara also illustrates a further significant deviation from the initial a priori definition of "surprise" relative to $H_1$. If the alternative $H_1$ is that this tombsite is that of the NT family, the "surprisingness" should indeed be assessed with respect to $H_1$ and not (only) with respect to the frequency table of the names. To illustrate this point consider a changed configuration of only the three male inscriptions, from (Yeshua son of Yoseph, Yoseh and Matya) to (Yoseh son of Matya, Jacob and Yoseph). Note that there is no Yeshua, and Yoseh is the son of an arbitrary Matya. Although a priori the changed configuration is by no means a serious candidate for being the NT family tombsite, under the suggested method the new configuration would have had a lower RR value than the actual one, that is, a higher "surprise."

3.2. *"Other" and disqualifying names.* Now let us address other features of the presumably a priori selected relevant lists. The relevant lists are allowed to include any number of names defined as "Other" as possibly belonging to persons about whom we have no records, with individual RR value of 1. Using this rule, the author computes the overall RR values as a product of the RR values of only four out of the six inscribed ossuaries (!). The ossuaries inscribed as Yehuda son of Yeshua (#2) and Matya (#3), although discussed at length, contribute nothing to the computation of the overall RR value. Following the rules set up by the suggested approach, this procedure is at least questionable. A set of rules which weigh positively (i.e., with a coefficient less than 1) names expected under $H_1$, but does not weigh negatively names which are unexpected under $H_1$, is likely to bias in favor of $H_1$.

Also, and continuing the previous point, it is mentioned that "...the list of persons (but not necessarily *names*) that would *disqualify* the tombsite as belonging to the NT family includes Joseph, Simon, and Yehuda" (as the persons' death did not occur in the relevant period of time, but the names may belong to other persons about whom we have no records). But if, say, an ossuary inscribed "Simon" would have been found in that tombsite (say, instead of that of "Matya") how could we have known whether it belongs to "that" Simon (brother of Jesus) or not? According to the "surprisingness" approach, we would have ignored that inscription, as belonging to "Other" (as belonging to a person about whom we have no records) and set the relevant coefficient to 1. The calculated $p$-value would have been exactly as in the present case. How can one thus judge the relevance to $H_1$ and render judgment about disqualifying? The overall impression is that the inevitable exposure to the data affected the definition of the provisos in favor of $H_1$.

**4. Another analysis.** I mentioned above that the inclusion of MM in the relevant list has a substantial effect on the overall results and conclusions. We can get an idea of the order of magnitude of that effect by comparing the results presented in Feurverger's paper with those yielded by another Bayesian analysis performed on the same data by Kilty and Elliot (2007). They consider the name Mariamene [η] Mara as irrelevant, and treated it identically to the names on the ossuaries inscribed Yehuda son of Yeshua, and Matya. Their computation is based on a listing of 32 scenarios of combinations of names one might expect to find in a NT family tombsite, based on Jesus' brothers and mother. All the scenarios have to include the Yeshua son of Yoseph (in any rendition), and are assumed to be equally probable. The a posteriori probability that this is indeed the tombsite of the NT family given the data is estimated by Kilty and Elliot as 0.487, very different from the values of well above 0.994, deduced from the odds ratios mentioned in Feuerverger's article.

The comparison between Kilty and Elliot's results and the a posteriori probabilities computed by Feuerverger illustrates the effect of the inclusion of Mariamene [$\eta$] Mara in Feuerverger's list. Obviously, other analyses of this data set are possible and indeed some are presented in articles posted on the internet. I refer to Kilty and Elliot's article, since unlike others, they mention that they agree in principle with Feuerverger's conclusions and their intention in writing the article was to show that the cluster of name is "hardly what a person should expect to find randomly." They further state that their figure is "quite comparable to Feuerverger's conclusion even though the two are done from very different standpoints." The statement seems to be inaccurate, probably based on fragmentary information of Feuerverger's results.

**5. Some final remarks.** Feuerverger emphasizes the provisos for the calculations, and mentions that the conclusion and the measure of surprisingness are based on a particular—but not uncontested—set of assumptions. He mentions that "as long as the definition of surprise is specified fully and a priori, the resulting approximate "tail area" will essentially be valid." It is difficult to accept that in this case, the elements of the new approach which are mentioned in the paper that have to be a priori specified (the hypothesis for the problem, the measure of surprisingness, the list of possible candidates, and the lists of nested possible name rendition for each candidate), have indeed been so specified. The final sentence in the paper candidly, and in my opinion very correctly, points to the weakest link in the foundation of the entire exposition and conclusions: "It is the presence in this burial cave of the ossuary of Mariamenou [$\eta$] Mara, and the mysteries concerning the identity of the woman known as Mary Magdalene, that hold the key for the degree to which statistical analysis will ultimately play a substantive role in determining whether or not the burial cave at East Talpiot happens to be that of the family of Jesus of Nazareth."

Let me re-phrase this sentence: "If the ossuary inscribed Mariamenou [$\eta$] Mara is indeed the ossuary of the Mary Magdalene from the New Testament, then, given the other names inscribed on the other ossuaries and the assumptions presented in the paper, we can state with a very high degree of confidence that that is the tombsite of the NT family."

I agree to such a statement. The only problem is that no statistical expertise is necessary to reach such a conclusion. If indeed, an ossuary *proven* to be that of Mary Magdalene was to be found, and in the same tombsite were also to be found ossuaries inscribed as Yeshua son of Yoseph, Yoseh and Marya, it is unlikely that the archeologists and the historians would appeal to statisticians for help. In such a case, as mentioned, the ossuary of the Mary Magdalene would have been by itself an important historical relic.

On the other hand, if we don't have that level of confidence regarding the Mary Magdalene ossuary, we have to rely on statistical analysis. Unfortunately, in my opinion, the stated principles of setting the assumptions were not followed, both in the presumably a priori compilation of the relevant lists as well as in the definition of the RR values (which allows discarding data which may point toward $H_0$ and assigns "surprisingness" values based the rareness of name frequencies rather than the actual closeness to $H_1$). The resulting effect on the conclusions reached is dramatic. Indeed, the narrator in the movie [Cameron (2007)] announced that Feuerverger's model concludes that "there is only one chance in 600 that the Talpiot tomb is not the Jesus family tomb, if Mary Magdalene can be linked to Mariamene." Later, in an interview on the *Scientific American* website [Mims (2007)], Feuerverger is quoted as saying that "I did permit the number one in 600 to be used in the film. I'm prepared to stand behind that but on the understanding that these numbers were calculated based on assumptions that *I was asked to use*," a statement far removed from the rigorous demand of a priori assumptions. [On his webpage, Feuerverger (2007) mentions that the quotations in the interview are "sufficiently accurate to be considered fair".]

In spite of the fact that, in my opinion, the analysis of the "surprisingness" based on the configuration of names failed to yield the stated conclusions, I refrain in this article from passing judgment on the subject matter issue of whether or not this is the tombsite of the NT family.

Furthermore, notwithstanding the reservations from the analyses applied to the discussed data, I applaud the bold initiative taken in the discussed paper to develop a new approach to tackle a problem characterized by a degree of complexity that precludes the straightforward application of the classical hypothesis framework. The general problem of rending judgment on whether a multiple characteristics observation represents the pursued specific entity or it is just the result from random draws is interesting and intriguing. Cases of disputed paternity and DNA matching come to mind in this context. Unlike the Talpiot case, in those cases a standard for comparison is available. The new approach and concepts of "surprisingness," "relevance" and "rareness" may evolve and prove beneficial in cases in which there is no such standard exists.

Classical methods, usually based on Bayesian analysis are available for those cases, but their application may be difficult in complex situations. If the new approach is to be applied, its performance needs to be compared to existing methods in situations in which it is known whether the null hypothesis (or the analogous null hypothesis) is correct. I think that the features of the approach still need to be investigated theoretically or by simulations under various conditions of complexity. In any case, the assumptions have to be pre-specified to ensure valid results and a valid comparison.

Department of Statistics and Operations Research
Raymond and Beverley Sackler Faculty of Exact Sciences
Tel Aviv University
Ramat Aviv, Tel Aviv
Israel 69978
E-mail: fuchs@post.tau.ac.il